\documentstyle[12pt]{article}
        \def\be{\begin{equation}}
        \def\ee{\end{equation}}

\begin{document}

\title {\normalsize \bf Algebraic and Geometric Structure of the Integrable Models \\ recently Proposed by Calogero
}

\author {V. Karimipour}
\begin{center}
\maketitle{\it Institute for Studies in Theoretical Physics and Mathematics\\
P. O. Box 19395-5531, Tehran , Iran \\ Department of Physics , Sharif
Uinversity of Technology \\P O Box 11365-9161, Tehran , Iran \\
} \end{center}  \vspace {12 mm}

\begin{abstract}
We show that the integrability of the dynamical system recently proposed by Calogero
and characterized by the Hamiltonian $ H = \sum_{j,k}^{N} p_j p_k \{\lambda +
\mu  cos [ \nu ( q_j - q_k)] \} $ is due to a simple algebraic structure
. It is shown that the integrals of motion are related to the Casimiar invariants of
of the $su(1,1)$ algebra. Our method shows clearly how these types of systems can be
generalized .
\end{abstract}
\newpage
\section{Introduction}
It has been recently shown [1,2] that the dynamical system characterized by the
Hamiltonian \be H = \sum_{j,k = 1}^{N} p_j p_k \{\lambda +  \mu  cos [ \nu ( q_j - q_k)] \}\ee
and the standard poisson brackets \be \{q_i , q_j \} =  \{p_i , p_j \} = 0 \ \ \ \
  \{q_i , p_j \} = \delta_{ij} \ee
is completely integrable.   Before going\  on\  we\  perform\  a canonical\  transformation $
q\longrightarrow {q\over \nu} \ \ \ \  p\longrightarrow { \nu p } $
and set the parameter $\nu $ equal to unity.
In refs.[1,2] Calogero showed that this system has the following properties:
\\
1) The quantities \be c_{jk} = p_j p_k \{ 1 - cos (q_j - q_k )\} \ee
are constants of the motion.\\
2) There are N independent constants of the motion in involution with each
other and H is among these integrals. These are the total momentum $ P = \sum_{i=1}^{N} p_i $  and :
\be h_m = \sum_{j,k=1}^{m} c_{j,k} \ \ \ \ m = 2, 3, . . N \ee
$$ \{h_m , h_{m'} \} = \{h_m , P \} = 0 $$
3) The quantities $C$ and $S$ defined by :
\be C = \sum_{j=1}^{N} p_j \  \cos  q_j \ \ \ \ \ \   S = \sum_{j=1}^{N} p_j \sin\   q_j    \ee
are constants of the motion if $ \lambda + \mu = 0 $  or if $ P = 0 $ , otherwise they evolve simply
in time.\\
4) The following extra relations hold:
\be \{ C , P \} = - S \ \ \ \  \{ S , P \} = C
\ \ \ \ \{ C , S \} = - P \ee  \be \{ C , H \} = - 2(\lambda + \mu ) PS \ \ \ \  \{ S , H \} =  2(\lambda + \mu ) PC \ee

The last two relations verify the statment made in part 3 above. And finally \\
5) The initial value problem for this Hamiltonian system was solved in explicit
form .\\

The starting point of Calogero and his main line of reasoning is to demand that
a Hamiltonian of the general form \be H = \sum_{j,k = 1}^{N} p_j p_k f(  q_j - q_k ) \ee
has constants of the motion of the following form \be c_{jk} = p_j p_k g ( q_j - q_k )
\ee  where $ f $ and $ g $ are
functions to be specifed. By this requirement he arrives at a functional equation
for $ f $  and $ g $ , one solution of which leads to the Hamiltonian (1) and the conserved
quantities (3). However the mutual poisson bracket of these integrals of motion
are complicated [1] \newpage
$$ \{ c_{jk} , c_{j'k'} \} = \delta_{jj'} p_j p_k p_k' \{ sin (q_j - q_k)
+ sin (q_k - q_k') + sin (q_{k'} - q_j)\} + $$ \be  ( j\leftrightarrow k )
+ ( j'\leftrightarrow k') + ( j\leftrightarrow k , j'\leftrightarrow k') \ee
The clever guess of Calogero is that the quantities given in (4)
are the required integrals of motion which are in involution ,hence the integrabil
ity of the system.  \\
He also demonstrated that the equations of motion can be derived from a Lax pair.

After all these calculations one is tempted to ask the following natural questions:
:\\
Is there any algebraic structure behind the integrability of this system?
Is the integrability of this system  related somehow to the existance of classical Yang Baxter
matrix or to some bi-Hamiltonian strucutre.? Can one
construct more general systems?
\\

It is the aim of this paper to answer the above questions.\\
We will find that the integrability of these systems is due to a very simple algebraic
and geometrical structure
which is related to the long range interactions and the factorizability
of the Hamiltonian.
These structure are completely different from the ones which are encountered
in the study of systems with local interactions.

\section{The Algebraic Structure}
Lets define the variables \be  x_j = p_j \ \cos q_j \ \ \ \ y_j = p_j \ \sin q_j \ \ \ \
 z_j = ip_j \ee

In the following we also use the notation $ x^1 = x \ \ \  x^2 = y \ \ \  x^3 = z $.
The Hamiltonian can now be written as:
\be H = \sum_{j,k = 1}^{N} - \lambda z_j z_k  +  \mu ( x_j x_k + y_j y_k ) \ee
where the new variables satisfy the follwoing $ su(2) $ poisson bracket relations:
\be \{ x_i^a , x_j^b \} = i \epsilon ^{abc} x_j^c \delta _{ij} \ee
{\bf Remark} : We use the complex number $i$ only for notational convenience in later manipulations
.In fact the poisson bracket between the real dynamical variables $ x_i \ \ y_i $ \ \ and
$ z'_j = p_j $ is related to the  $ su(1,1) $ algebra.
\\
Now define the variables \be  X^a_m = x^a_1 + x^a_2 + x^a_3 + . . .    x^a_m \ee
It is obvious that for each $m$ these sets of variables satisfy the same relations
among themselves as in (13) and form a copy of $ su(2) $ algebra , and furthermore since the smaller copies of the algebra
are embeded in the larger copies we have:
\be \{ X_i^a , X_j^b \} = i \epsilon ^{abc} X^c_{(i,j)}  \ee
where $ (i,j) $ is meant to denote the minimum of $i$ and $j$ i.e:
\be \{ X_2^a , X_2^b \} = i \epsilon ^{abc} X^c_{2}
\ \ \ \ \ \ \{ X_2^a , X_3^b \} = i \epsilon ^{abc} X^c_{2}  \ee

Defining for each copy ,say the $m$-th one the Casimir function \be C_m = \sum_{a=1}^3 X_m^a X_m^a \ee
we obtain:
\be \{ C_i , X_j^b \} = 2i \epsilon ^{abc} X^a _i X^c_{(i,j)}  \ee
\be \{ C_i , C_j \} = 4i \epsilon ^{abc} X_i^a X_j^b X^c_{(i,j)}   \ee
We now note that in the last formula the indices $ i $ and $ j $ are not dummy variables
,however the index $ (i,j) $ is either equal to $i$
or to $j$ , in any case the tensor which is contracted with $\epsilon ^{abc}$
is symmetric with respect to the interchange of two of the indices $ (a,c) $ or $ (b,c) $ , hence the right hand side  identically vanishes:
$$\{C_i , C_j \} = 0 $$
It is interesting to note that although the Casimir of one copy  does not commute with the generators
of another copy as seen from (17) ,the Casimirs of different
copies commute among themselves.
The Hamiltonain can now be written as:
\be H =  -\lambda Z_N^2 + \mu ( X_N^2 + Y_N^2 ) =- ( \lambda  + \mu ) Z_N^2 +  C_N   \ee
Using (17,14,15) the following relations can also be verified directly:
\be \{ C_i , X_N^b \} = 0  \ \ \ \ \ b = 1, 2, 3 \ \ \ \ \ \ \{ H , Z_N \} = 0 \ \ \ \  \ee
\be \{ X_N , H \} = 2i ( \lambda + \mu ) Z_N Y_N  \ \ \ \ \ \ \ \{ Y_N , H \} =  - 2i ( \lambda + \mu ) Z_N X_N \ee
We see very clearly the essence of integrability of the system and have been able to avoid
the intermediate constants of the motion $ c_{jk}$ with their complicated poisson
brackets and directly reach the integrals of motion which are in involution.\\
{\bf Proposition}: The Casimiar functions $ C_i , i = 2 , 3 , . . .  N $ and $ P\equiv
-iZ_N  $ are N inetegrals of motion in involution with each other .
 ( Note that $ C_1 $ is identically equal to zero ) .The Casimir functions
$ C_m $ mudulo a minus sign are exactly equal to the integrals $ h_m $ defined in [1].\\
In fact  we have:\be C_m = \sum_{a=1}^{3} X_m^a X_m^a = \sum_{a=1}^{3} \sum_{j,k =1}^{m} x_j^a x_k^a = \ee
\be = \sum_{j,k =1}^{m} ( x_j x_k + y_j y_k + z_j z_k ) =  - \sum_{j,k =1}^{m} p_j p_k \{ 1 - cos (q_j - q_k )\} \ee
From (11) and (13) we readily find the algebraic meaning of all the quantities introduced in [1]
and summarized in the introduction:
\be Z_N = iP \ \ \ \ C_m = - h_m \ \ \ \  X_N = C \ \ \ \ Y_N = S \ee

All the algebraic and poisson bracket relations between these quantities found in
[1], follow from the above identification.
\section{Generalizations}
We can now generalize our construction and include the models of Calogero as a special case.
Let $ g $ be a simple lie algebra of rank $ r $  with generators $ e^a \ \ \ a = 1, 2 . . . dim\ g $\ \
and relations \be [ e^a , e^b ] = C^{ab}\ _c e^c \ee
And let $ g^* $ be its dual with basis $ e_a $ . It is well known [3] that the lie structure on $ g $ induces a poisson structre on $ C ( g^* ) $
\be \{ x^a , x^b \} = C^{ab}\ _c x^c \ee
where the $ x^a \ \ \ a = 1, 2, . . dim g = dim g^* $ are the local coordinates in $ g^* $ \\
In general the poisson bracket is degenerate , to make it nondegenerate one restricts it to the
submanifolds of $ C(g^*)$ naturally defined by setting the values of the Casimir functions equal to
constants. These submanifolds are always even dimensional and the poisson bracket
becomes symplectic on them.  By Darboux theorem one can then define local canonical coordinates and momenta
on these submanifolds. ( the analogue of eq. (11) above ). \\
One can now define the variables $ X^a_m $ exactly as in (13) . All the formalism
of section (2) can be followed exactly except that there  are $r$ Casimir functions
involved labeled by $$ C^{\alpha}_m \ \ \ \ \alpha = 1, 2, . . . r \ \ \ \ m =  2, 3, . . . N $$
with \be   \{ C^{\alpha}_m , C^{\beta}_n \} = 0 \ee
Furthermore one has \be \{ C^{\alpha}_m ,  H^{\beta}_N \} = 0 \ee
where $ H^{\beta}_N  $ correspond to the Cartan subalgebra elements of the N-th copy  of $ g$ .
Obviously any Hamiltonian of the  general form
\be H = H ( C^1_2, , , , C^{r}_N , H^1_N , .... H^r_N ) \ee
defines an integrable system which is a generalization of the one introduced in [1,2].

As a concrete application consider again the algebra $ su(1,1) $with relations
\be \{ x , y \} = - z' \ \ \ \ \{ y , z' \} = x  \ \ \ \ \{ z' , x \} = y  \ee
The symplectic submanifold are defined by  $$ x^2 + y^2 - z'^2  = c $$
where $ c $ is a constant. These submanifolds are of two completely different
geometry. For $ c = 0 $ they are double cones with a singularity at the apex.
This is the leaf chosen by Calogero with the canonical coordinates (11).
The other leaves where $ c\ne 0 $ are hyperboloids with the following
Canonical coordinates:
\be z' = p \ \ \ \ \ x = {\sqrt {p^2 + c}} \ \ \cos q \ \ \ \ \  y = {\sqrt {p^2 + c}} \ \ \sin q \ee
Hence a generalization of the model of [1] is defined by:
\be H = \sum_{j,k = 1}^{N} \lambda p_j p_k + \mu \{ {\sqrt {p^2_j + c}}\ \ {\sqrt {p^2_k + c}}\ \ \cos [ ( q_j - q_k)] \}\ee
with the integrals of motion given by :
$$h_m =\sum_{j,k = 1}^{m}  p_j p_k - \{{\sqrt {p^2_j + c}}\ \ {\sqrt {p^2_k + c}}\ \ \cos [ ( q_j - q_k)] \}  $$
and $$ P = \sum_{i=1}^{N} p_i $$

\section{ Discussion}
In addition to sheding light on the nature of integrability in this system
the algebraic approach proposed in this letter has several further consecquences:
\\ a) It shows how one can construct more general systems by using higher rank
algebras .
\\ b) With little modification it proves the integrability of similar systems at
the quantum level [4]
\\ b) By using the $su(2) $ algebra instead of the $su(1,1)$ one can prove the integrability of
classical  and  quantum Heisenberg  xxz magnets with long range interactions even in the presence of .
magnetic field [4].
\section{Acknowledgments} I would like to thank A. Aghamohammadi and S. Shariati for
interesting comments.
\newpage
{\large \bf References}
\begin{enumerate}
\item  F. Calogero : Phys. Lett. A. {\bf 201} (1995) 306-310
\item  F. Calogero : Jour. Math. Phys. {\bf 36} 9 (1995)
\item  L. D. Faddeev ; Integrable Models in 1+1 dimensional quantum Fields
theory
(Les Houches Lectures 1982), Elsevier, Amsterdam (1984)
\item V. Karimipour " Integrability of classical and quantum Heisenberg xxz magnet
with long range interactions in a magnetic field " In preparation :
\end{enumerate}
\end {document}